# Feed-forward and noise-tolerant detection of feature homogeneity in spiking networks with a latency code.

Michael Schmuker[1,2,*], Rüdiger Kupper[1],
Ad Aertsen[3], Thomas Wachtler[4], and Marc-Oliver Gewaltig[1,5]

1: Honda Research Institute Europe GmbH
2: University of Hertfordshire, Department of Computer Science, Biocomputation Group
3: Bernstein-Center Freiburg and Faculty of Biology, University of Freiburg
4: Ludwig-Maximilians-Universität München, Department of Biology II
5: École Polytechnique Fédérale de Lausanne, Blue Brain Project

* Corresponding author: m.schmuker@herts.ac.uk.

## Abstract

In studies of the visual system as well as in computer vision, the focus is often on contrast edges. However, the primate visual system contains a large number of cells that are insensitive to spatial contrast and, instead, respond to uniform homogeneous illumination of their visual field. The purpose of this information remains unclear. Here, we propose a mechanism that detects feature homogeneity in visual areas, based on latency coding and spike time coincidence, in a purely feed-forward and therefore rapid manner. We demonstrate how homogeneity information can interact with information on contrast edges to potentially support rapid image segmentation. Furthermore, we analyze how neuronal crosstalk (*noise*) affects the mechanism's performance. We show that the detrimental effects of crosstalk can be partly mitigated through delayed feed-forward inhibition that shapes bi-phasic post-synaptic events. The delay of the feed-forward inhibition allows effectively controlling the size of the temporal integration window and, thereby, the coincidence threshold. The proposed model is based on single-spike latency codes in a purely feed-forward architecture that supports low-latency processing, making it an attractive scheme of computation in spiking neuronal networks where rapid responses and low spike counts are desired.

## 1 Introduction

Humans recognize the presence of a face in an image in as little as 150 ms, which strongly suggests that object recognition in humans must be achieved in a feed-forward manner, with at most 1-2 spikes per neuron per processing stage (Thorpe and Imbert, 1989). Mechanisms for object recognition in the brain that involve recurrent processing for segmentation (Körner et al., 1999; Ullman, 2000) are inherently too slow to account for this kind of rapid face detection.

Gewaltig et al. (Gewaltig et al., 2003) proposed a mechanism that supports rapid image segmentation by using information on homogeneous image regions to confine edge detection to borders between those regions. Here, we present a straightforward spiking network model that implements this effect. The application of this model is not limited to vision but could be applied universally to detect homogeneous distributions of features in any spiking sensory code.

The model was inspired by the response properties of a neuron population in the primate visual pathway, the so-called konio-cellular neurons. In primates, the lateral geniculate nucleus (LGN) relays visual information from the retina to the primary visual cortex. Three parallel pathways relay information through the LGN: The magno-, parvo-, and koniocellular pathways. Magno-cells in the LGN have large receptive fields and are highly sensitive to local contrast (Sclar et al., 1990). Parvo-cells have smaller receptive fields and their contrast sensitivity is comparatively low, however, many of them are sensitive to color (Livingstone and Hubel, 1988). The koniocellular pathway is less well investigated. Konio-cellular neurons form a heterogeneous group, intercalated between the magno- and parvo layers in the LGN. Many of them have response properties distinct from those found in magno- and parvocellular neurons, with large and non-classical receptive fields with lower spatial and intermediate temporal resolution compared to parvo- and magno-cells at matched eccentricities, without centre-surround organization, and/or insensitive to flashes or moving bars (Irvin et al., 1986; White et al., 2001; Xu et al., 2001). Different types of K-cells with "blue-on" and "blue-off" characteristics have been described, and even cells suppressed by contrast (Szmajda et al., 2006; Tailby et al., 2007; Solomon et al., 2010). These observations indicate that at least a subset of the koniocellular pathway transmits information about homogeneous illumination in large receptive fields, rather than local contrast patches as in the magnocellular and parvocellular pathways.

Subgroups of Konio-cells also show unusual projection patterns to striate cortex. Their projections to V1 largely avoid the magno- and parvocellular-targeted Layer 4, but instead terminate in the cytochrome-oxidase blobs of the superficial layers (Hendry and Yoshioka, 1994; Lund et al., 1994; Hendry and Reid, 2000) and directly influence activity in supragranular layers (Klein et al., 2016). Moreover, while magno- and parvo-cellular projections are restricted to V1, konio-cellular inputs have been identified in area MT, a higher area of the ventral visual pathway (Sincich et al., 2004; Jayakumar et al., 2013). These extrastriate projections have been associated with "blindsight" in subjects that have V1 ablated but are still able to perceive rudimentary visual information (Hernández-González et al., 1994).

These unique response profiles and projection patterns provoke the question what role konio-cells play in vision. Here, we propose that information about homogeneous regions of the visual scene, as provided by a subset of konio-cells, can be used to filter bottom-up information on contrast edges. The particular projection pattern from the LGN directly to Layers 2/3 suggests that konio-cell axons "skip" a synapse and can interact with contrast-edge information provided by magno-and parvo-cellular neurons entering V1 in Layer 4, before being relayed to Layers 2/3. Filtering bottom-up information on contrast edges could thereby suppress the output of contrast-sensitive cells in Layers 2/3 in regions with homogeneous brightness, potentially supporting rapid image segmentation.

Rapid processing of information in spiking network has also been associated with latency coding. There is convincing experimental (Reinagel and Reid, 2000; Johansson and Birznieks, 2004; Nelken et al., 2005; Gollisch and Meister, 2008) and theoretical (Reinagel and Reid, 2000; Van Rullen and Thorpe, 2001; Delorme, 2003; Kupper et al., 2005; VanRullen et al., 2005; Masquelier and Thorpe, 2007) evidence that the spike latency of neurons encodes their stimulation strength, such that the stronger a neuron is activated, the sooner it produces a spike. Latency can readily be assessed with the first spike of a neuron, usually after a few milliseconds, whereas the read-out of spike rates requires integrating many spikes over time or space. Latency codes are thus preferred when there is a need to quickly arrive at a hypothesis about the stimulus (VanRullen et al., 2005). The model we present here combines spike latency coding of with coincidence detection to identify regions of homogeneous illumination that could assist rapid feed-forward segmentation.

In the visual system stimuli are generally mapped topographically, such that feature selective neurons projecting to a single post-synaptic neuron have their receptive fields located in a small local region of the input space (e.g., a small local region in the field of view, Fig. 1, *left*). Assuming those neurons respond to visual input with a latency code, the level of coincidence of the first spikes after the common reference point (i.e., stimulus onset, termination of a saccade) will encode the amount of variance in the input patch. In the case of low variance, i.e. *homogeneous* stimulation, the ensemble will produce a spike packet within a small temporal window. This enables post-synaptic neurons to detect homogeneous stimulation by coincidence detection (Fig. 1, *right*). Therefore, a spike in the post-synaptic neuron becomes a statement on the homogeneity of the distribution of the respective feature in this region of the input space.

By definition, a latency code requires a reference point in time. In the visual system of mammals, this reference point could be provided e.g. by the onset and termination of saccades (Ito et al., 2011; Niemeyer and Paradiso, 2018). Such "reset" mechanisms synchronize the internal state of an ensemble of neurons (e.g. the membrane potential), thereby erasing their stimulus history, which is a necessary condition for reliable mapping of local luminances on the spike latencies in a population of neurons (Kupper et al., 2005).

The above example applies to vision and luminance-encoding neurons. But this principle of detecting feature homogeneity by spike coincidence applies to any topographically arranged set of neurons that use a latency code to represent the strength of their activation. It can easily be extended to retinotopically arranged orientation-selective cells in V1, or retinotopically arranged directionally sensitive neurons in MT. Moreover, it is not limited to vision – any sensory modality that supports latency coding and some sort of topographical mapping could be processed in the proposed fashion.

# 2 Models and Methods

To design a generic homogeneity-detecting circuit we started from a set of neurons selective for some arbitrary feature *f*, with the aim to create a set of cells selective for the *homogeneous appearance of f*. We required that feature-selective neurons are topographically arranged over the (potentially abstract) input space, and that this topography is preserved in convergent projections across layers.

The design pattern for the neural detection of topographical feature homogeneity was chosen to be as follows (cf. Fig. 2):

1. A sending neuronal population $N_s$ is a set of latency-coding neurons selective for the feature *f*. That is, spiking of a neuron in $N_s$ corresponds to the appearance of feature *f*, and the spike latency corresponds to some gradual quality $q$ of *f* (e.g., illumination, contrast, orientation). The gradual quality of $q$ allows it to be mapped to the spike latency.
2. The neuronal population $N_s$ is arranged across the cortical surface, while preserving the topography of a stimulation space $\mathbf{T}$, e.g. retinotopically, tonotopically, somatotopically, chemotopically etc.
3. Neurons in a receiving population $N_r$ receive convergent, topography-preserving inputs from local sub-populations of $N_s$ with a fixed extent (or, in the case of visual receptive fields, diameter) $d$ measured in $\mathbf{T}$, and with a fixed transmission delay. This makes $N_r$ topographically arranged according to $\mathbf{T}$.

Then, activity of a neuron in population $N_r$ indicates the appearance of feature $f$ with homogeneous quality $q$ across a local region of diameter $d$ in the stimulation space $\mathbf{T}$, and at the location corresponding to the respective neuron's topographical position. Note that the quality $q$ of feature $f$ is still encoded in the firing latency of this neuron.

We applied this general design principle in two network models: First, a model of V1 where homogeneity selective cells identified surfaces in natural images, and second, a generalized model where we analyzed the effect of neuronal crosstalk (or, colloquially, *noise*) on the performance of homogeneity detection. We show that surface-detection in the V1 model can support stimulus processing in the primate visual cortex. The model consists of two retinotopically arranged sets of spiking neurons in feedforward fashion (Fig. 2). The sending network stage $N_s$ models a retinotopic set of LGN-konio-cells that encode local luminance to spike latency. The receiving network stage $N_r$ applies spike-latency based homogeneity detection to the output of $N_s$, making $N_r$ selective for image regions of homogeneous luminance (surfaces). Since regions of homogeneous luminance and contrast edges excluded each other, the output of surface-selective neurons can be used to suppress unsought responses in orientation selective cells, which do not define borders between mid-sized visual objects (Gewaltig et al., 2003).

In the generalized model, we focused on the coding properties of our latency-based principle for homogeneity detection. We examined the robustness to neuronal crosstalk at levels to be expected in the brain.

We used the NEST simulator developed in collaboration with the Neural Simulation Technology Initiative (Gewaltig and Diesmann, 2007) for simulating the structured neuronal network. The simulation had a temporal resolution of 0.1 ms.

## 2.1 A model of cooperative feature detection in V1

Our first model is a specific application of the design pattern described above. We modeled LGN and V1 responses for the konio-cellular and the parvo- or magnocellular pathways. For this model, we assume for mathematical conveniency two classes of konio-like cells that respond to homogeneous illumination of their receptive field, responding to bright (ON) and dark (OFF) stimuli. We further assume a population of coincidence detecting neurons in V1 that these konio-like cells project to.

The schematic model setup is shown in Fig. 3. It consisted of four retinotopic network stages:

1. a retinal input layer where natural gray-scale images were applied,
2. a layer of cells in the LGN that responded to homogeneous illumination, as in a subpopulation of the *konio-like* pathway ($N_s$), consisting of ON- and OFF-cells,
3. a layer of homogeneity-selective coincidence detecting neurons in V1, driven by the konio-like LGN population ($N_r$, ON and OFF),
4. a layer of parvo- or magno-cellular orientation-selective cells in V1 (four orientations).

### 2.1.1 LGN cells responding to homogeneous illumination (*konio-like*) ($N_s$)

We applied a two-step process to model the responses of *konio-like* LGN cells responding to homogeneous illumination.

First, we convolved the input image $I(x,y)$ with a Gaussian kernel $G(x,y)$, mimicking the effect of a large receptive field, and subsequently passed the result through a sigmoidal activation function $\Theta$ (Gewaltig et al., 2003):

$$A \coloneqq \Theta(I * G) \tag{1}$$

where

$$G\colon\ (x,y) \mapsto \frac{1}{\sigma\sqrt{2\pi}} exp\left(-\frac{x^2+y^2}{2\sigma^2}\right) \tag{2}$$

$$\Theta\colon\ z \mapsto \frac{1}{1+exp(-2b(z-\theta))} \tag{3}$$

with $\sigma$ being the spatial standard deviation of the Gaussian, and $\theta, b$ the threshold and slope of the sigmoid. The threshold $\theta$ was chosen according to the mean of pixel values in the image. The other parameters were arbitrarily set to $b = 4$ and $\sigma = 2$ for illustration purposes.

Second, the activation values $A(x, y)$ were transformed into spike trains by a leaky integrate-and fire model. We used a current based leaky integrate-and-fire neuron model (Lapicque, 1907; Tuckwell, 1988) with post-synaptic currents shaped as $\alpha$-functions. Their membrane potential dynamics are given by

$$\tau_m \frac{dV}{dt} = E_l - V(t) + R_m I(t) , \qquad (4)$$

where $\tau_m$ is the membrane time constant (set to 10 $ms$), $R_m$ the membrane resistance (40 $m\Omega$), and $E_l$ the leakage, i.e., resting potential ($-70$ $mV$). An action potential was produced when the membrane potential $V$ crossed the firing threshold $V_{th}$ ($-55$ $mV$) in upgoing direction. Each neuron was allowed to fire only one spike.

We transformed the activation values of konio-like LGN cells into neuronal spike responses by injecting currents of corresponding magnitude into a retinotopically arranged layer of these model neurons. An ON- and an OFF-cell was located at each retinotopic position. The interval of activation values $[0, a_{max}]$ was linearly mapped to an interval of injection currents $[I_0, I_1]$:

$$[0, a_{max}] \mapsto [I_0, I_1] \qquad (5)$$

$$a \to I_0 + (I_1 - I_0) \frac{a}{a_{max}} \qquad (6)$$

with

$$[I_0, I_1] = \begin{cases} [450,750]\ pA \text{ for ON cells} , \\ [750,450]\ pA \text{ for OFF cells.} \end{cases} \qquad (7)$$

The time of the first spike in the leaky integrate-and-fire neuron model after the onset of a constant injection current $I$ was taken to be:

$$t_{cross} = -\tau_m \ln\left(\frac{R_m I + E_l - V_{th}}{R_m I + E_l - V_{start}}\right), \qquad (8)$$

provided that the current was strong enough to drive the neuron to cross threshold (Dayan and Abbott, 2001). $V_{start}$ was the membrane potential at which the neuron started integrating the stimulus current. We will later consider its effect on the neuronal latency code. Here, we assumed that the neurons had relaxed to their resting membrane potential at the time of stimulus onset ($V_{start} = E_l$). Thus, Equation (8) becomes

$$t_{cross} = -\tau_m \ln\left(1 - \frac{V_{th} - E_l}{R_m I}\right) . \qquad (9)$$

The benefit of ON- and OFF-cells becomes apparent in Fig. 4, which depicts the non-linear relation between the injection current $I$ and the first-spike latency $t_{cross}$. The injection current is a function of pixel luminance (Eqs. 1-3) and represents the post-retinal activation. The mapping between current and spike latency is logarithmic (*cf.* Eq. 9). Consequently, the temporal spread of a spike packet created from an image patch will depend on the overall brightness of the patch. For example, consider a dark patch that results in injection currents between 450 and 500 $pA$. ON cells responding to that image patch will fire their spikes within a 4 ms wide window (Fig. 4). The same image patch, at higher brightness (600 to 650 $pA$), yields a spike packet only 1 ms wide. The homogeneity information may be identical in both patches, yet the bright patch might elicit a spike in coincidence-detecting surface-selective cells in $N_r$, whereas the dark patch might fail to do so, because the temporal spread of the spike packet is too large.

Hence the problem arises that surface-selective neurons would rarely respond to homogeneous dark image regions with homogeneity thresholds that yielded good results for bright image regions. Likewise, tuning the model to detect dark homogeneous regions by loosening the coincidence threshold would lead to spurious responses to non-homogeneous bright regions.

Using two classes of LGN-konio-cells (ON and OFF) with inverse characteristics for post-retinal activation (Eq. 7) mitigated the most extreme effects of this nonlinearity (although it did not completely eliminate them). ON- and OFF-cells projected separately to two corresponding classes of surface-selective V1 neurons: V1-ON-cells, the firing threshold of which was set to accurately detect bright homogeneous regions, i.e., mapping bright pixel luminance to a large range of spike latencies, and V1-OFF-cells, with the same firing threshold but inverse luminance-to-current characteristics, and therefore accurately detecting dark homogeneous image regions. Parameters were adjusted such that both ON- and OFF-detectors still performed reasonably well in areas of medium luminance.

#### 2.1.2 V1 homogeneity-selective neurons ($N_r$)

The konio-like LGN acted as the sending population $N_s$ in our coincidence-based feature detection scheme (Fig. 2). The V1 neurons in our network model corresponded to konio-driven neurons in the cytochrome-oxidase (CO) blobs of the superficial layers 2/3 in the primary visual cortex. We modeled the konio-driven V1 neurons as a retinotopic layer of leaky integrate-and-fire model neurons, playing the role of the receiving population $N_r$ in our design pattern for feature homogeneity detection (Fig. 2). Konio-driven V1 neurons received spike input from locally restricted populations of luminance-selective LGN-konio-cells, as shown in Fig. 2.

We used the same neuron model as for the LGN cells, with $\alpha$-shaped excitatory postsynaptic currents with a time constant of $\tau_{syn} = 0.63\ ms$. Receptive field size was $5 \times 5$ pixels with uniform weights (0.4, unitless), except for the four pixels in the corners, which had zero weight to approximate a circular receptive field. We used a very short membrane time constant of $\tau_m = 30\ \mu s$ to obtain a point neuron model that is sensitive to spike coincidence, rather than rate input.

#### 2.1.3 V1 orientation-selective neurons

We modeled the response of parvo- or magno-cellular orientation-selective cells in V1 in a two-step process, similar to the one used for the LGN-konio-cells:

1. We computed activation values for four classes of orientation-selective neurons by convolving the input image with Gabor-filters of four orientations, spaced at 45-degree intervals.
2. The activation values were then transformed into spike trains by current-injection into a leaky integrate-and-fire model, as described before for LGN cells.

At each position of the retinotopic layer, four orientation-selective neurons were located (preferring contrast orientations of 0, 45, 90, 135 degrees, respectively).

#### 2.1.4 Cooperation between orientation- and homogeneity-selective features in V1

In the initial phase of visual scene interpretation, the information provided by the surface-selective neurons can potentially be helpful in a quick segmentation of the visual input. To obtain a suitable hypothesis on the main contents of the scene "at a first glance", the rich information in the stimulus needs to be reduced to its most salient parts.

Orientation-selective and homogeneity-selective neurons convey information that is mutually exclusive, as the presence of an oriented contrast edge contradicts the fact that the receptive field is illuminated homogeneously. Suppressing responses of orientation-selective cells in areas where homogeneity-selective neurons with co-located receptive fields are active could therefore support rapid feed-forward segmentation (Gewaltig et al., 2003).

Such feed-forward interaction between the K-pathway and the M/P pathways is supported by anatomical and physiological evidence. Many K-neurons project directly to the CO blobs in V1, skipping a synapse in layer 4, compared to M and P cells (Ding and Casagrande, 1998). The temporal properties of K cells are otherwise comparable to those of M and P cells (Solomon et al., 1999). Nevertheless, electric field potentials evoked in human V1 by S-cone isolating stimuli appear earlier (latencies of about 40 ms) than the common luminance-defined motion-specific potentials (Morand et al., 2000). This indicates that konio-cellular responses in V1 are present at least as early as the responses of orientation-selective neurons. Since edges and surfaces of physical objects exclude each other, the output of surface-selective konio-neurons can be used to suppress the responses in orientation-selective cells that are might be less useful in this initial phase of scene analysis, i.e., those orientation responses which do not define borders between visual objects (Gewaltig et al., 2003). We modeled this process by feedforward inhibitory coupling from surface-selective konio-cells onto the retinotopically corresponding orientation-selective parvo- or magno-cells in V1.

## 2.2 Generalized model

In the following we use a generalized network architecture similar to the V1 model (Fig. 2), albeit with a simpler stimulation scheme. We dropped the OFF-channel because it is functionally identical to the ON-channel, and we dropped the low-pass filtering and contrast enhancement stages that were used to mimic the konio-cellular properties of the primate LGN.

To mimic and control the influence of neuronal cross-talk, each neuron in $N_r$ received input from two pools of Poissonian background noise (Fig. 2), the excitatory pool $np1$ with 16,000 neurons firing independently at 2 spikes/s, and an inhibitory pool $np2$ with 4,000 neurons firing at 0.787 spikes/s. The rates and weights of the two noise pools were calibrated using a “balanced neuron” paradigm, such that the model neuron spiked at a mean rate of 2 spikes/s background activity per neuron, i.e. as if it were a member of the noise pool. Synaptic currents were set to 15 pA for excitatory and 150 pA for inhibitory synapses. These values are well within the physiological range (see e.g. (Thomson and Lamy, 2007)). Taken together, each neuron in the $N_r$ population received background input from 20,000 independent neurons, maintaining them at a spontaneous rate of 2 spikes/s in the absence of external stimuli.

### 2.2.1 Stimulation

We used a $100 \times 100$ pixels patch from a gray-scale image from an in-house database, representing a section of $10 \times 10$ degrees in the visual field. Pixel luminance $l$ was linearly mapped to an injection current $I$ without preprocessing:

$$[black, white] \hat{=} [0,1] \mapsto [I_0, I_1] \quad (10)$$

$$l \mapsto I_0 + l(I_1 - I_0) \quad (11)$$

with

$$[I_0, I_1] = [376,800]\, pA\,. \quad (12)$$

The currents were injected into a layer of 100×100 leaky integrate-and-fire neurons ($N_s$). The spatial mapping of image pixels to stimulated neurons was retinotopic. Luminance was mapped to a slightly different current range compared sec. 2.1, that starts at the minimum current that will elicit a spike. The absolute value of the range was not relevant to the general principle we intended to illustrate.

### 2.2.2 Neuronal reset

The relation between injected current $I$ and the time of the first spike after stimulus onset is described by (Eq. 8). Previously we assumed the membrane potential $V_{start}$ at which the neuron starts integrating the stimulus current to be fixed. However, it is obvious that the crossing time $t_{cross}$, i.e., the neuron's spike latency, will vary if $V_{start}$ varies, depending on the individual neurons' stimulation history. These variations can be quite dramatic, as illustrated by Fig. 5. Small variations in $V_{start}$ cause strong variations in $t_{cross}$, especially in the physiologically relevant excitation regime below 600 $pA$ (corresponding to 100 spikes/s of tonic firing, i.e., a latency of 10 $ms$ starting from resting potential; Fig. 5, thick curve). A reset mechanism that ensures identical starting membrane potentials is therefore required to generate a coherent latency code across the $N_s$ population. In V1, such a reset which erases a neuron's stimulus history can be generated by saccades (Ito et al., 2011; Niemeyer and Paradiso, 2018). In our model, we reset the neurons' membrane potentials to resting potential at the beginning of the stimulus presentation, which was meant to represent the termination of a saccade. Visual stimulation was applied constantly thereafter for 100 $ms$.

This procedure made the $N_s$ population a retinotopically arranged set of latency-coding neurons, selective for local luminance (luminance, in this example, was the feature $f$).

### 2.2.3 Homogeneity-selective cells

Sub-populations of $N_s$ in circular regions of diameter $d = 11$ neurons (representing $d = 1.1$ degree visual angle) projected in a convolutional, topography-preserving fashion onto the $100 \times 100$ $N_r$ population (Fig. 2). Transmission delays and synaptic weights were identical for all connections. Hence, $N_r$ is a retinotopically arranged set of neurons, selective for spatially homogeneous luminance across distances of $d = 1.1$ degree visual angle. Fig. 9A illustrates receptive field sizes for $N_s$ and $N_r$ neurons.

# 3 Results

## 3.1 Cooperative homogeneity detection in a model of V1

The simulated homogeneity-selective LGN cells formed a retinotopically arranged set of latency-coding feature selective neurons. The firing latency of these neurons represented the average luminance in their large receptive fields.

We applied grey-scale images from an in-house database to our model retina and recorded the spike responses of all model neurons. From these data we derived topographically arranged *response maps* for the different layers of model neurons.

Since LGN neurons encode the mean luminance in their receptive fields, V1 neurons will fire if their receptive field is illuminated homogeneously. One could argue that regions of homogeneous luminance can often be attributed to the surfaces of mid- to large-sized objects, and, hence, describe the konio-driven V1 neurons as *surface-selective* (Gewaltig et al., 2003). These V1 neurons correspond to targets of the koniocellular pathway that are found in V1 layer 2/3. Since parvo- and magnocellular pathways usually terminate in layer 4 and project to layer 2/3 from there, the koniocellular inputs in layer 2/3 are "one synapse ahead". This provides the potential for homogeneity-selective neurons to suppress bottom-up input from contrast-edge sensitive parvo-/magnocellular neurons before it arrives in layer 2/3.

Fig. 6 illustrates the effect of inhibiting edge-detector responses in regions where model konio-driven cells respond to homogeneous illumination. The input image in Fig. 6a is processed by orientation/edge detectors (Fig. 6b) and surface detectors (Fig. 6c). Inhibiting responses of orientation-selective cells in homogeneous image regions yields a qualitative segmentation of the image that emphasizes responses of edge detectors between objects and suppresses them within object boundaries. It should be noted that Fig. 6c is an overlay of ON- and OFF-responses (Fig. 7).

The input neurons representing konio-cellular LGN neurons were modelled to have large receptive fields compared to the edge detectors. Switching off the low-pass filtering resulted in detecting homogeneity on a much finer spatial scale that appears less efficient in suppressing fine-grained image details (Fig. 8). This highlights the requirement for large receptive fields for surface-processing, i.e., to direct visual processing to large and salient objects in the initial phase of scene analysis, such that a first hypothesis on scene content can be established.

In summary, we demonstrated that the combination of neurons encoding homogeneous illumination of their receptive field in spike latency and coincidence-based detection of homogeneous areas (*surfaces*) in the visual field can restrict responses of orientation-selective cells, with the potential to support rapid feedforward image segmentation.

### 3.2 Generalized model and noise resistance

Representation of mutually exclusive features in two parallel pathways for rapid processing is not restricted to vision or the substrate of V1 but could act generally to detect the homogeneous appearance of (potentially abstract) latency-encoded features in any sensory modality and the cortical areas.

So far, we did not consider crosstalk from other neuronal processes that are unrelated to the “signal” in question. Among a neuron’s ten thousand synapses (Braitenberg and Schüz, 1991) the ones transmitting the “signal” (spikes from a local set of feature detectors) may make up one percent or less. The majority of post-synaptic potentials will therefore be caused by spike trains that are unrelated to the “signal” in question. Such activity is often called “background activity”, “ongoing activity”, or simply “noise” in the neuroscience literature (e.g. (Arieli et al., 1996)). It does not imply that these inputs are useless noise that could be simply removed.

Therefore, in the following, we tested the noise-resistance of our mechanism for homogeneity detection. Specifically, we investigated whether its operation degrades gradually (*gracefully*) with increasing noise level.

We recorded spike responses of all neurons over the complete time-course of simulation. From this data, we retrieved topographical latency maps of $N_s$ and $N_r$ neurons, as well as topographical maps of response probability. For each stimulus presentation we reset the $N_s$ membrane potentials and applied a static image for stimulation, generating latency codes for the applied stimulus. The response latency of a neuron was defined as the time span between the reset and the neuron’s first spike thereafter.

Over the course of the entire simulation, the $N_r$ neurons received synaptic input from the two noise pools that imitated neuronal background activity. This massive bombardment with spikes from the excitatory and inhibitory noise pools drove the $N_r$ neurons close to firing threshold and caused them to fire at a spontaneous rate of 2Hz in the absence of a visual stimulus, due to fluctuations in the membrane potential (the so-called “*fluctuation-driven regime*”; e.g. (Kuhn et al., 2004)). To test the noise resistance of our network model, we conducted all experiments with varying strengths of input from the background noise pools, by scaling the synaptic weights from the noise pools to 0% (no noise), 50%, and 100%.

Fig. 9B shows a raster plot of typical spike trains generated by $N_s$ neurons along the horizontal line in Fig. 9A, relative to the last reset time at $t = 0$. Regions of homogeneous luminance in the stimulus patch show up as “spike packets”, since groups of neighboring $N_s$ neurons respond with similar latencies (Fig. 9B). Fig. 10A, left column, shows corresponding typical spike-trains from $N_r$ in the absence of noise. They are driven by coincident spikes in $N_s$, enabling an ad-hoc segmentation of the input into quasi-homogeneous regions of similar (high) luminance, separated by regions where no spikes occur (Fig. 10A, right column). $N_r$ neurons thus act as *homogeneity-selective neurons*. Note that by relying only on the first action potentials of feature detector neurons, homogeneity processing was very fast, with first components signaled already after $10\,ms$. Increasing noise levels of 50% and 100% (Fig. 10B and C, left column) progressively jitter $t_{cross}$ across $N_r$ neurons, leading to degraded segmentation (right column).

Fig. 11 shows the responses of all 100 $N_r$ neurons across 100 repeated reset and stimulation cycles. Spike generation in $N_r$ is deterministic in the absence of noise (Fig. 11A). As noise is added, $N_r$ neurons produce spurious spikes, even in dark image regions where $N_s$ neurons are silent, and in inhomogeneous regions where incoming $N_s$ spikes do not form a coincident packet (Fig. 11B&C). Fig. 11A-C, right column displays spike probability histograms. The strictly bimodal distribution of spike probability in the absence of noise becomes spread out as the noise level increases to 50%, although spike probabilities were still low in non-homogeneous regions. At 100% noise all stimulus regions had a spike probability above 0.4 and the histogram approached a unimodal distribution. Under such circumstances homogeneous and non-homogeneous regions become effectively indistinguishable. This demonstrates that the utility of a pure latency code to identify homogeneous input regions would be limited under biologically realistic noise levels of 2 spikes/s per neuron, without further mechanisms to mitigate noise.

Another observation from comparing Figs. 9 and 10 is that the latency of $N_r$ neurons depended on the average latency of the original spike packet from $N_s$. In other words, $N_r$ neurons were *transparent* to the latency code of the afferent neurons, encoding how strongly activated this feature was. They blocked spike responses from non-homogeneous regions and preserved the latency code from homogeneous regions.

## 3.3 Shaping postsynaptic potentials by feedforward inhibition

Adding noise to the input of $N_r$ neurons naturally introduces jitter in their response times to otherwise unchanged stimuli. This jitter adds to the variance observed in inputs from upstream $N_s$ neurons. Assuming we cannot influence the background noise, to make the mechanism more robust against noise we must address the jitter in the input.

Since $N_r$ neurons are coincidence detectors, incoming $N_s$ spikes must overlap within the synaptic integration time window. This time window depends on two factors, the membrane time constant and the synaptic time constant. We used the $\alpha$-function to model synaptic currents (Fig. 12, black curve) in a leaky integrate-and-fire neuron model ((Lapicque, 1907; Tuckwell, 1988); see also (Kuhn et al., 2004)). In our model, the synaptic current $I(t)$ had a relatively short rise time of $2\,ms$, followed by a longer "excitatory tail" that kept having a positive (depolarizing) contribution to the membrane potential for at least $10\,ms$. Roughly, in order for synaptic currents to add up and drive the postsynaptic neuron above threshold, they must coincide in this $12\,ms$ window.

Generally, spikes arriving at random times (such as from background noise) are less likely to add up when the time window for integration is short. The impact of background noise can therefore be mitigated by shortening the synaptic time constant. This can be achieved either through varying the model parameters for membrane resistance and capacity, or synaptic time constant directly. In biological terms this equates to having a specialized neuron type for coincidence detection. While there are numerous examples for specialized neuron and synapse types in the brain (e.g. the "calyx of Held" synapses of neurons in the medial nucleus of the trapezoid body, that maximize transmission probability and minimize latency (Borst and Soria Van Hoeve, 2012)), such extreme specialization would not support generic neuronal computation in large cortical neural ensembles where each neuron may participate in different types of computation at different points in time. Computational studies have shown that membrane time constants are affected by ongoing background activity (Kuhn et al., 2004), but purposefully controlling synaptic integration via this mechanism may, however, be difficult.

Bi-phasic post-synaptic potentials generated by feed-forward inhibition provide an alternative way of altering the effective synaptic integration window, without the need to tweak neuronal parameters. Fig. 12 shows the effect of an inhibitory post-synaptic current (PSC) delivered shortly after an excitatory PSC of the same magnitude and time constants. The currents superimpose and result in a much narrower effective PSC. Moreover, the net charge delivered by the paired pulse is zero, preventing the postsynaptic potential from accumulating over a longer time. Both effects raise the temporal constraints for constructive superposition of subsequent synaptic events. For more detailed accounts of the working of feedforward inhibition and its effect on spike timing accuracy, including its application in a V1 network model, we refer to (Kremkow et al., 2010b, 2010a, 2016) and the references mentioned therein.

To include PSC shaping by feedforward inhibition into our model, we duplicated the connections from the sending $N_s$ layer to the receiving $N_r$ layer by adding feedforward inhibitory projections with a fixed delay. Fig. 13 shows the spike probabilities of the homogeneity-detecting $N_r$ neurons for a range of inhibitory delays. The total area classified as homogeneous, as indicated by high $N_r$ spike probability, increased with increasing inhibition delay. Short delays lead to shorter effective PSCs, and incoming spike packets would need to be more focused in time in order to elicit postsynaptic spikes. Thereby, the delay of the paired feedforward inhibition effectively determined the threshold, at which a local region was classified as homogeneous by the $N_r$ neurons.

Shorter effective PSCs not only affected the homogeneity threshold, they also enabled homogeneity detection under conditions of strong background noise, where this would otherwise not have been possible. Returning to Fig. 11C, a realistic level of neuronal crosstalk rendered homogeneity detection infeasible. Fig. 11D shows the same setting with applying shaped PSCs by feedforward inhibition. The inhibition delay was 8 $ms$, the same as in the upper right panel of Fig. 13. Biphasic synaptic events helped to improve separating positive from negative detector responses, as the spike probability histogram in Fig. 11$D$ was clearly bimodal, implying that homogeneity detectors tended to either respond with a high or a low probability to the stimulus. The separation of homogeneous from inhomogeneous image parts in full-strength noise was then comparable to the case in Fig. 11$B$, with 50% noise strength, effectively mitigating the impact of crosstalk.

# 4 Discussion

We presented a spiking neuronal network motif for detecting homogeneous stimulus features. In a set of neuronal feature detectors (visual or other modality), it naturally extracts local populations of neurons that are equally stimulated relative to a common reference point in time. Since spike latency is a monotonic function of stimulation strength, equally stimulated pre-synaptic neurons fire coincident spike waves, which are easily detected in a receiving neuron. The mechanism exploits that post-synaptic currents caused by action potentials are temporally restricted events. A coincidence detector neuron is brought above firing threshold by many of those events temporally superimposing.

The model is restricted to one spike per neuron per response to a stimulus, in line with Thorpe's postulate that psychophysically confirmed timing constraints for human visual object detection in humans can only be achieved with a maximum of one or two spikes per processing stage (Thorpe and Imbert, 1989). As a useful side effect, this approach avoids spurious coincidences that might otherwise occur in high-spike-rate processes. The one-spike-per-neuron rule could potentially be relaxed to entail short bursts of a few spikes, which might be a more realistic representation of the true neurobiological process.

We motivated the homogeneity detector concept from properties of a subset of cells in the koniocellular pathway in the primate visual system. It should be noted that the koniocellular pathway is composed of a very heterogeneous group of cells and that only a subset of them has large receptive fields that respond to a homogeneous illumination without antagonistic surround (Irvin et al., 1986; White et al., 2001; Xu et al., 2001; Szmajda et al., 2006; Tailby et al., 2007; Solomon et al., 2010). There is strong physiological and anatomical evidence for the specific wiring of these cells, targeting layers 2/3. Much recent work in visual neuroscience involved rodents, rather than primates. It has been hypothesized that certain groups of cells in the rodent visual system may be homologous to cells in the K-pathway, but to our knowledge the evidence supporting such homology is scarce and inconclusive (Denman and Contreras, 2016). Rodents and primates occupy very different ecological niches, and the evolutionary pressure on their visual systems is likely to prioritize different modes of function. Therefore, we base our model exclusively on biological evidence collected in primates.

The homogeneity detector circuit we propose responds to areas of homogeneous illumination, that can assist rapid segmentation of the visual scene by interacting with edge-detector responses. Homogeneously stimulated regions in sensory space will often constitute relevant stimuli, like uniformly colored objects, uniform texture, uniform movement. The simple circuitry needed to implement it (local convergence) makes it a biologically plausible candidate for all sensory cortices.

To be applicable in a dense and complex neuronal system like the brain, a neuronal mechanism must be robust to noise and crosstalk from millions of neighboring and, possibly interfering, neurons. We observed that our mechanism was not robust enough to base a reliable decision upon. At a level auf crosstalk that is to be expected in the brain, homogeneity responses were basically indistinguishable from noise baseline (Fig. 11*C*).

To improve performance, we introduced forward inhibition, which created biphasic synaptic currents with short excitatory parts, and zero net current into the receiving neuron. Both effects decreased the sensitivity to neuronal crosstalk and allowed for better separation of homogeneous stimuli. At a realistic level of neuronal crosstalk, correct homogeneity responses were substantially more probable than false events (bimodal probability distribution in Fig. 11*D*). This means that reliable decisions could be based on the pooling of homogeneity responses, even in a noisy brain.

The principle of PSC shaping opens up an attractive opportunity for dynamically controlling the mode of operation in a spike-coding network. Information on sensory stimuli is encoded in both firing rate and temporal structure of spike trains (Tsodyks and Markram, 1997; Reinagel and Reid, 2000; Van Rullen and Thorpe, 2001; Nelken et al., 2005). Dynamic transition from latency to rate processing in the very same neurons can unify the processing of spike-time and rate codes in a single network, and hence forms an attractive scheme of computation in neural networks. As demonstrated, PSC shaping can be used to control the overlap of synaptic events in the post-synaptic neuron. It can potentially be used to select which of the two components of information a postsynaptic neuron extracts, controllable by changing feedforward inhibition with no change in network structure, and no change in processing of the pre-synaptic neurons.

It also gives rise to potential homeostatic mechanisms that could adjust the effective PSC on the basis of the fraction of active cells in the population. Separation of homogeneous from inhomogeneous image regions could be achieved by setting a threshold, e.g. at 50% of spiking cells. In the active brain, the delay at which inhibitory action potentials are produced can be influenced by network effects. Diffuse background stimulation may change the effective firing threshold of neurons (Chance et al., 2002). The neurophysiological effects of selective attention can promote neuronal processing (Hochstein and Ahissar, 2002; Müller and Kleinschmidt, 2004). The latency code can be read out in a flexible way: The same mechanism that improves robustness to noise (feedforward inhibition) can also dynamically change the threshold at which homogeneity is detected. Hence, it is conceivable that the homogeneity threshold may dynamically change, according to the momentary cognitive requirements or interests.

Besides features like color, orientation, etc. (Itti and Koch, 2000), homogeneity might constitute an important component of a visual saliency map. Since contrast edges and homogeneous surfaces of physical objects are mutually exclusive, the output of surface-selective neurons can be used to suppress responses in orientation-selective cells in regions with homogeneous luminance, supporting the initial phase of scene analysis (Gewaltig et al., 2003). We demonstrated in our model of V1 that this function can be implemented in a simple feed-forward network of parallel feature-extraction through parvo-/magno- and konio-cellular pathways.

Parallel extraction of anticorrelated and therefore possibly redundant features incurs additional metabolic cost that needs to be outweighed by the gain in reliability and speed of processing. A detailed analysis of this cost is outside the scope of this study. One factor mitigating the additional cost is that homogeneity-selective neurons have much larger receptive fields than contrast-edge detectors and require less dense packing. Another is that feedforward processing in parallel pathways is the only way to determine different aspects of an instantaneous stimulus in cases where lateral interaction is impossible, as suggested by response time analyses in the visual system (Thorpe et al., 1996).

Our model focused on detecting homogeneous luminance in a retinal image. But the mechanism could be applied to any other feature that can be expressed in a latency code, at any phase in processing, be it visual or in other modalities, like auditory or somatosensory. For example, it is conceivable to use the coincidence detection mechanism to detect repetitive patterns in the input, i.e. homogeneous texture. If $f$ is a feature that is spatio-temporally defined, spatially homogeneous changes or spatially homogeneous motion can potentially be detected.

Generalizations of the proposed mechanism may also find applications outside biology for neuromorphic signal processing and pattern recognition when combined with recent results in spiking networks and image processing. For example, a generalized approach to the information-filtering properties of populations of coincidence detector neurons has been suggested (see e.g. (Bostner et al., 2020) and references therein). Further, the concept of time-surfaces has been introduced for recognition of moving objects and correlated stimuli in event-based vision (Lagorce et al., 2017). Time-surfaces represent a spatio-temporally coherent change of feature detector unit outputs that may be detectable by coincidence detector neurons. Finally, a recently proposed method for image quantization demonstrates a practical application of the concept (Christie et al., 2020): Homogeneous image areas are detected by their low gradient energy, and subsequently encoded in lower bit depth than regions with high gradient intensity, reducing average image bit depth and enabling lower-power image processing. These examples provide an interesting outlook for future research not just in theoretical brain research, but also towards practical utility of this mechanism for signal processing.

# 5 Acknowledgments

We thank Dr. Ursula Körner for her help and support in compiling the evidence on koniocellular processing strategies across primate species, that was instrumental in forming the hypothesis underlying this study. This work was supported by BMBF grant 01GQ0420 to BCCN Freiburg, and EU H2020 grant 945539 (Human Brain Project SGA3) to the University of Hertfordshire.

# 6 References

Arieli, A., Sterkin, A., Grinvald, A., and Aertsen, A. (1996). Dynamics of ongoing activity: explanation of the large variability in evoked cortical responses. *Science (80-. ).* 273, 1868–1871.

Borst, J. G. G., and Soria Van Hoeve, J. (2012). The calyx of held synapse: From model synapse to auditory relay. *Annu. Rev. Physiol.* 74, 199–224. doi:10.1146/annurev-physiol-020911-153236.

Bostner, Ž., Knoll, G., and Lindner, B. (2020). Information filtering by coincidence detection of synchronous population output: analytical approaches to the coherence function of a two-stage neural system. *Biol. Cybern.* 114, 403–418. doi:10.1007/s00422-020-00838-6.

Braitenberg, V., and Schüz, A. (1991). *Anatomy of the Cortex: Statistics and Geometry*. Berlin, Heidelberg, New York: Springer Verlag.

Chance, F. S., Abbott, L. F., and Reyes, A. D. (2002). Gain modulation from background synaptic input. *Neuron* 35, 773–782. doi:10.1016/S0896-6273(02)00820-6.

Christie, O., Rego, J., and Jayasuria, S. (2020). Analyzing Sensor Quantization of raw images for visual SLAM. in *Proceedings of ICIP 2020*, 2484.

Dayan, P., and Abbott, L. F. (2001). *Theoretical neuroscience: Computational and mathematical modeling of neural systems*. Cambridge MA, London England: MIT Press Available at: http://mitpress.mit.edu/book-home.tcl?isbn=0262041995.

Delorme, A. (2003). Early cortical orientation selectivity: How fast inhibition decodes the order of spike latencies. *J. Comput. Neurosci.* 15, 357–365. doi:10.1023/A:1027420012134.

Denman, D. J., and Contreras, D. (2016). On Parallel Streams through the Mouse Dorsal Lateral Geniculate Nucleus. *Front. Neural Circuits*. doi:10.3389/fncir.2016.00020.

Ding, Y., and Casagrande, V. A. (1998). Synaptic and neurochemical characterization of parallel pathways to the cytochrome oxidase blobs of primate visual cortex. *J. Comp. Neurol.* 391, 429–443. doi:10.1002/(SICI)1096-9861(19980222)391:4<429::AID-CNE2>3.0.CO;2-2.

Gewaltig, M.-O., and Diesmann, M. (2007). NEST (NEural Simulation Tool). *Scholarpedia* 2, 1430.

Gewaltig, M. O., Körner, U., and Körner, E. (2003). A model of surface detection and orientation tuning in primate visual cortex. *Neurocomputing* 52–54, 519–524. doi:10.1016/S0925-2312(02)00767-1.

Gollisch, T., and Meister, M. (2008). Rapid Neural Coding in the Retina with Relative Spike Latencies. *Science (80-. ).* 319, 1108–1111. doi:10.1126/science.1149639.

Hendry, S. H. C., and Reid, R. C. (2000). The Koniocellular Pathway in Primate Vision. *Annu. Rev. Neurosci.* 23, 127–153. doi:10.1146/annurev.neuro.23.1.127.

Hendry, S. H. C., and Yoshioka, T. (1994). A neurochemically distinct third channel in the macaque dorsal lateral geniculate nucleus. *Science (80-. ).* 264, 575–577. doi:10.1126/science.8160015.

Hernández-González, A., Cavada, C., and Reinoso-Suárez, F. (1994). The lateral geniculate nucleus projects to the inferior temporal cortex in the macaque monkey. *Neuroreport* 5, 2693–2696. doi:10.1097/00001756-199412000-00071.

Hochstein, S., and Ahissar, M. (2002). View from the top: Hierarchies and reverse hierarchies in

the visual system. *Neuron* 36, 791–804. doi:10.1016/S0896-6273(02)01091-7.
Irvin, G. E., Norton, T. T., Sesma, M. A., and Casagrande, V. A. (1986). W-like response properties of intralaminar zone cells in the lateral geniculate nucleus of a primate (Galago crassicaudatus). *Brain Res.* 362, 254–270.
Ito, J., Maldonado, P., Singer, W., and Grun, S. (2011). Saccade-Related Modulations of Neuronal Excitability Support Synchrony of Visually Elicited Spikes. *Cereb. Cortex* 21, 2482–2497. doi:10.1093/cercor/bhr020.
Itti, L., and Koch, C. (2000). A saliency-based search mechanism for overt and covert shifts of visual attention. *Vision Res.* 40, 1489–1506. doi:10.1016/S0042-6989(99)00163-7.
Jayakumar, J., Roy, S., Dreher, B., Martin, P. R., and Vidyasagar, T. R. (2013). Multiple pathways carry signals from short-wavelength-sensitive ('blue') cones to the middle temporal area of the macaque. *J. Physiol.* 591, 339–352. doi:10.1113/jphysiol.2012.241117.
Johansson, R. S., and Birznieks, I. (2004). First spikes in ensembles of human tactile afferents code complex spatial fingertip events. *Nat. Neurosci.* 7, 170–177. doi:10.1038/nn1177.
Klein, C., Evrard, H. C., Shapcott, K. A., Haverkamp, S., Logothetis, N. K., and Schmid, M. C. (2016). Cell-Targeted Optogenetics and Electrical Microstimulation Reveal the Primate Koniocellular Projection to Supra-granular Visual Cortex. *Neuron* 90, 143–151. doi:10.1016/j.neuron.2016.02.036.
Körner, E., Gewaltig, M.-O., Körner, U., Richter, A., and Rodemann, T. (1999). A model of computation in neocortical architecture. *Neural Networks* 12, 989–1005. doi:10.1016/S0893-6080(99)00049-0.
Kremkow, J., Aertsen, A., and Kumar, A. (2010a). Gating of signal propagation in spiking neural networks by balanced and correlated excitation and inhibition. *J. Neurosci.* Available at: http://www.jneurosci.org/content/30/47/15760.short [Accessed January 30, 2014].
Kremkow, J., Perrinet, L. U., Masson, G. S., and Aertsen, A. (2010b). Functional consequences of correlated excitatory and inhibitory conductances in cortical networks. *J. Comput. Neurosci.* 28, 579–94. doi:10.1007/s10827-010-0240-9.
Kremkow, J., Perrinet, L. U., Monier, C., Alonso, J.-M., Aertsen, A., Frégnac, Y., et al. (2016). Push-Pull Receptive Field Organization and Synaptic Depression: Mechanisms for Reliably Encoding Naturalistic Stimuli in V1. *Front. Neural Circuits* 10, 37. doi:10.3389/fncir.2016.00037.
Kuhn, A., Aertsen, A., and Rotter, S. (2004). Neuronal Integration of Synaptic Input in the Fluctuation-Driven Regime. *J. Neurosci.* 24, 2345–2356. doi:10.1523/JNEUROSCI.3349-03.2004.
Kupper, R., Gewaltig, M.-O., Körner, U., and Körner, E. (2005). Spike-latency codes and the effect of saccades. *Neurocomputing* 65–66, 189–194. doi:10.1016/j.neucom.2004.10.006.
Lagorce, X., Orchard, G., Galluppi, F., Shi, B. E., and Benosman, R. B. (2017). HOTS: A Hierarchy of Event-Based Time-Surfaces for Pattern Recognition. *IEEE Trans. Pattern Anal. Mach. Intell.* 39, 1346–1359. doi:10.1109/TPAMI.2016.2574707.
Lapicque, L. (1907). Recherches quantitatives sur l'excitation electrique des nerfs traitee comme une polarization. *J. Physiol. Pathol. générale* 9, 620–635.
Livingstone, M., and Hubel, D. (1988). Segregation of form, color, movement, and depth: Anatomy, physiology, and perception. *Science (80-. ).* 240, 740–749. doi:10.1126/science.3283936.
Lund, J. S., Yoshioka, T., and Levitt, J. B. (1994). "Substrates for Interlaminar Connections in Area V1 of Macaque Monkey Cerebral Cortex," in *Cerebral Cortex: Primary Visual Cortex in Primates*, eds. A. Peters and K. S. Rockland (Plenum Press), 37–60. doi:10.1007/978-1-4757-9628-5_2.
Masquelier, T., and Thorpe, S. J. (2007). Unsupervised learning of visual features through spike timing dependent plasticity. *PLoS Comput. Biol.* 3, e31. doi:10.1371/journal.pcbi.0030031.
Morand, S., Thut, G., De Peralta, R. G., Clarke, S., Khateb, A., Landis, T., et al. (2000). Electrophysiological evidence for fast visual processing through the human koniocellular pathway when stimuli move. *Cereb. Cortex* 10, 817–825. doi:10.1093/cercor/10.8.817.
Müller, N. G., and Kleinschmidt, A. (2004). The attentional 'spotlight's' penumbra: center-surround modulation in striate cortex. *Neuroreport* 15, 977–980. doi:10.1097/00001756-200404290-

00009.
Nelken, I., Chechik, G., Mrsic-Flogel, T. D., King, A. J., and Schnupp, J. W. H. (2005). Encoding Stimulus Information by Spike Numbers and Mean Response Time in Primary Auditory Cortex. *J. Comput. Neurosci.* 19, 199–221. doi:10.1007/s10827-005-1739-3.
Niemeyer, J. E., and Paradiso, M. A. (2018). Saccade-based termination responses in macaque V1 and visual perception. *Vis. Neurosci.* 35, E025. doi:10.1017/S0952523818000032.
Reinagel, P., and Reid, R. C. (2000). Temporal Coding of Visual Information in the Thalamus. *J. Neurosci.* 20, 5392–5400. doi:10.1523/JNEUROSCI.20-14-05392.2000.
Sclar, G., Maunsell, J. H. R., and Lennie, P. (1990). Coding of image contrast in central visual pathways of the macaque monkey. *Vision Res.* 30, 1–10. doi:10.1016/0042-6989(90)90123-3.
Sincich, L. C., Park, K. F., Wohlgemuth, M. J., and Horton, J. C. (2004). Bypassing V1: a direct geniculate input to area MT. *Nat. Neurosci.* 7, 1123–8. doi:10.1038/nn1318.
Solomon, S. G., Tailby, C., Cheong, S. K., and Camp, A. J. (2010). Linear and nonlinear contributions to the visual sensitivity of neurons in primate lateral geniculate nucleus. *J. Neurophysiol.* doi:10.1152/jn.01118.2009.
Solomon, S. G., White, A. J., and Martin, P. R. (1999). Temporal contrast sensitivity in the lateral geniculate nucleus of a New World monkey, the marmoset Callithrix jacchus. *J. Physiol.* 517 ( Pt 3, 907–17. doi:10.1111/j.1469-7793.1999.0907s.x.
Szmajda, B. A., Buzás, P., FitzGibbon, T., and Martin, P. R. (2006). Geniculocortical relay of blue-off signals in the primate visual system. *Proc. Natl. Acad. Sci. U. S. A.* doi:10.1073/pnas.0606970103.
Tailby, C., Solomon, S. G., Dhruv, N. T., Majaj, N. J., Sokol, S. H., and Lennie, P. (2007). A new code for contrast in the primate visual pathway. *J. Neurosci.* doi:10.1523/JNEUROSCI.5343-06.2007.
Thomson, A. M., and Lamy, C. (2007). Functional maps of neocortical local circuitry. *Front. Neurosci.* 1, 19–42. doi:10.3389/neuro.01.1.1.002.2007.
Thorpe, S., Fize, D., and Marlot, C. (1996). Speed of processing in the human visual system. *Nature* 381, 520–522. doi:10.1038/381520a0.
Thorpe, S. J., and Imbert, M. (1989). “Biological constraints on connectionist models,” in *Connectionism in perspective*, eds. R. Pfeifer and F. Fogelman-Soulié (Amsterdam: Elsevier), 63–92.
Tsodyks, M. V., and Markram, H. (1997). The neural code between neocortical pyramidal neurons depends on neurotransmitter release probability. *Proc. Natl. Acad. Sci.* 94, 719–723. doi:10.1073/pnas.94.2.719.
Tuckwell, H. C. (1988). *Introduction to Theoretical Neurobiology*. Cambridge University Press doi:10.1017/cbo9780511623271.
Ullman, S. (2000). *High-Level Vision: Object Recognition and Visual Cognition*. Cambridge MA: MIT Press.
Van Rullen, R., and Thorpe, S. J. (2001). Rate coding versus temporal order coding: What the retinal ganglion cells tell the visual cortex. *Neural Comput.* 13, 1255–1283. doi:10.1162/08997660152002852.
VanRullen, R., Guyonneau, R., and Thorpe, S. J. (2005). Spike times make sense. *Trends Neurosci.* 28, 1–4. doi:10.1016/j.tins.2004.10.010.
White, A. J. R., Solomon, S. G., and Martin, P. R. (2001). Spatial properties of koniocellular cells in the lateral geniculate nucleus of the marmoset Callithrix jacchus. *J. Physiol.* 533, 519–535. doi:10.1111/j.1469-7793.2001.0519a.x.
Xu, X., Ichida, J. M., Allison, J. D., Boyd, J. D., Bonds, A. B., and Casagrande, V. A. (2001). A comparison of koniocellular, magnocellular and parvocellular receptive field properties in the lateral geniculate nucleus of the owl monkey (Aotus trivirgatus). *J. Physiol.* 531, 203–18. doi:10.1111/j.1469-7793.2001.0203j.x.

## Figures and legends

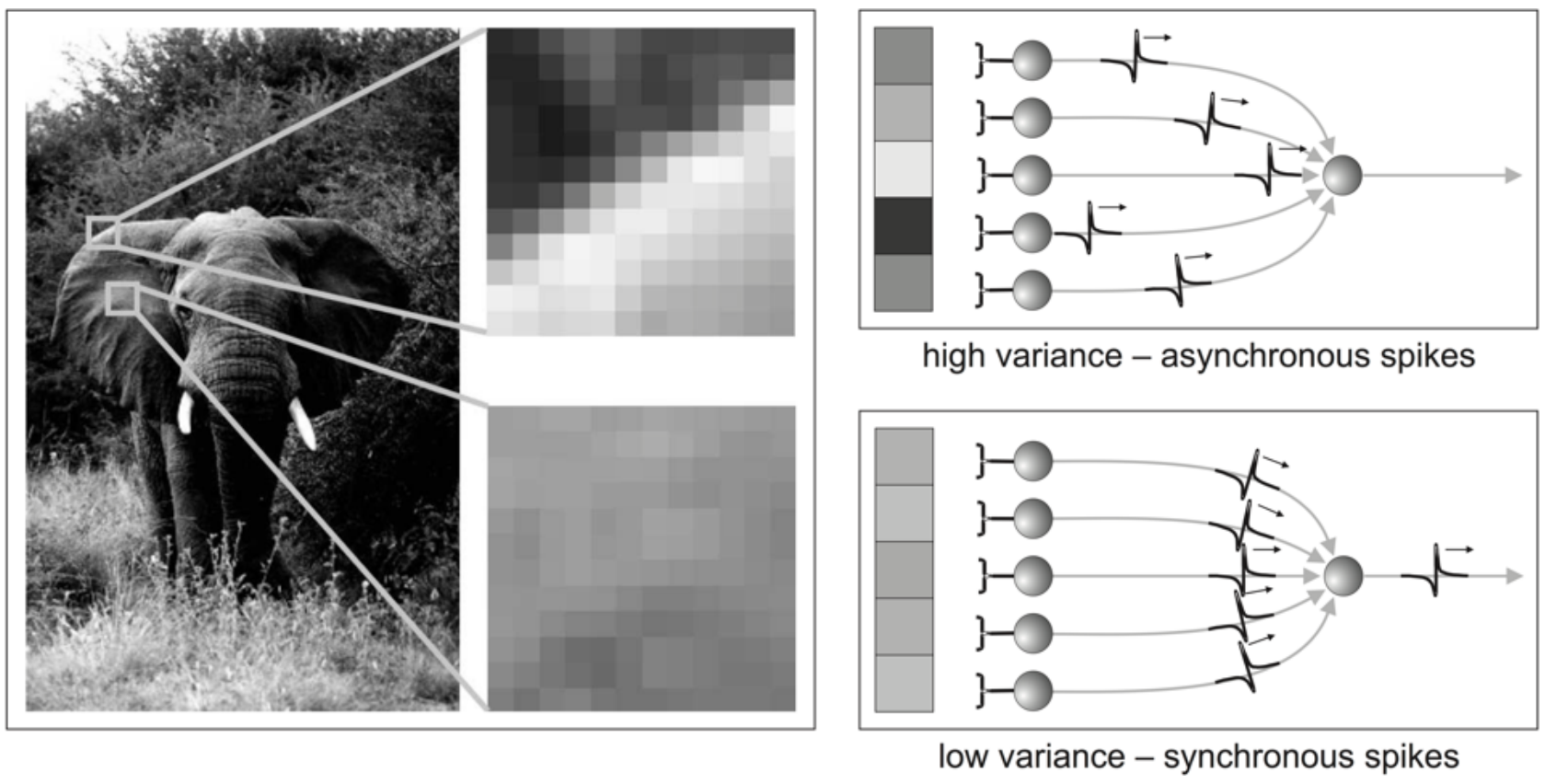

**Figure 1:** The principle of spike-latency based homogeneity processing; example from the visual domain. *Left:* Edge regions in the stimulus typically have high variance of luminance, whereas surface regions typically have low variance of luminance. *Right:* A set of visual neurons respond at latencies depending on the luminance inside their receptive fields. The receiving neuron responds when the incoming action potentials coincide. This makes it selective for homogeneous luminance inside its receptive field. In the visual domain, this allows identifying surface regions.

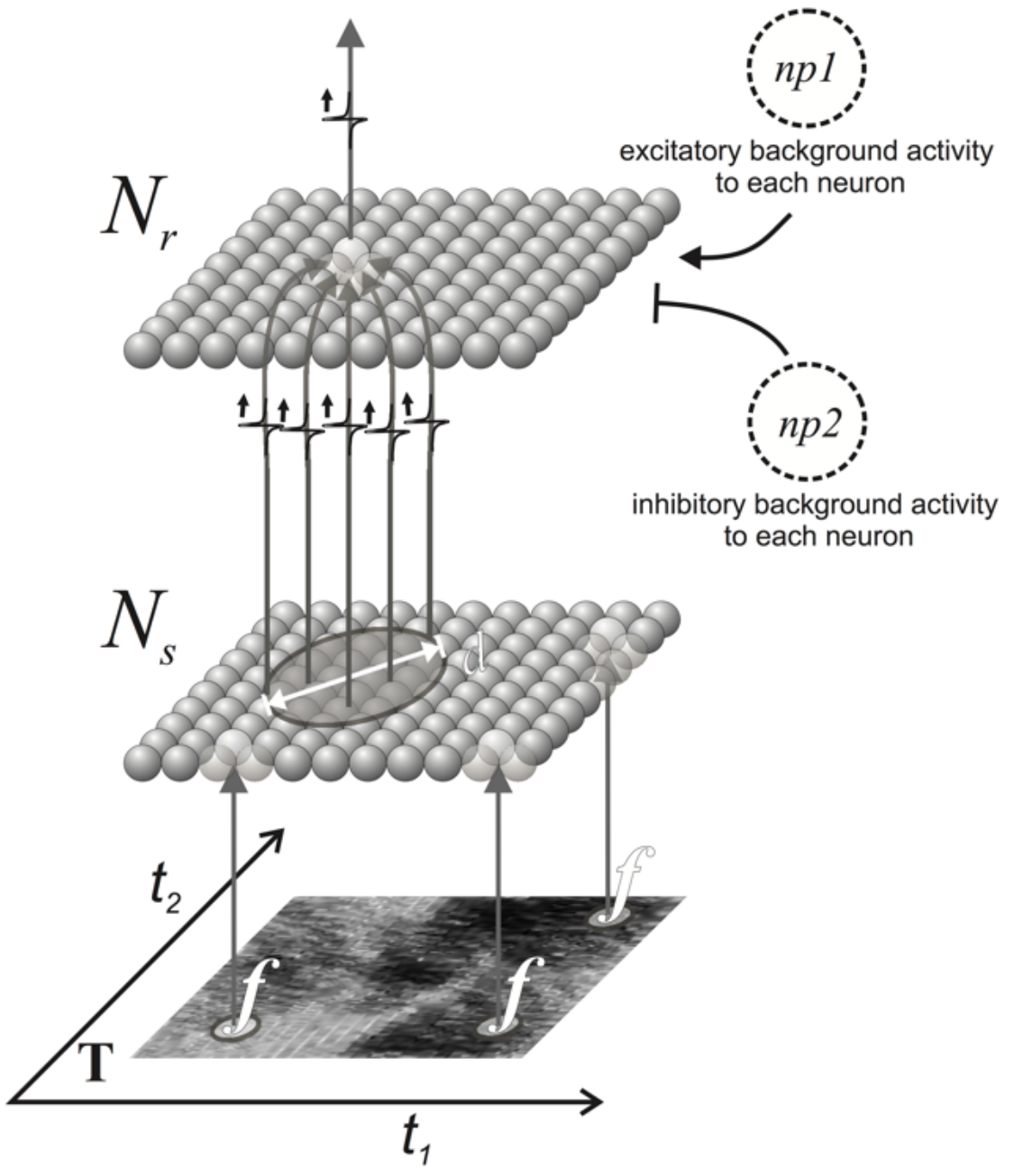

**Figure 2:** Detecting feature homogeneity from spike latency in a topographical arrangement of feature selective neurons. Homogeneity is detected in two network stages. $N_s$ is a topographically arranged set of neurons, selective for feature *f*. They generate a spike-latency code for the quality of feature *f* at the respective location in the input space **T**. A single $N_r$ neuron receives action potentials from a local patch of $N_s$ neurons. Its sensitivity for coincident synaptic events makes this neuron selective for the homogeneous appearance of feature *f* inside its receptive field. All $N_r$ neurons receive action potentials from topographically corresponding local patches of $N_s$. This makes $N_r$ a topographically arranged set of neurons, selective for the homogeneous appearance of feature *f*. In our generalized model (Sec. 3), two noise pools *np1* and *np2* provided random spikes, mimicking crosstalk from 20,000 unrelated neurons.

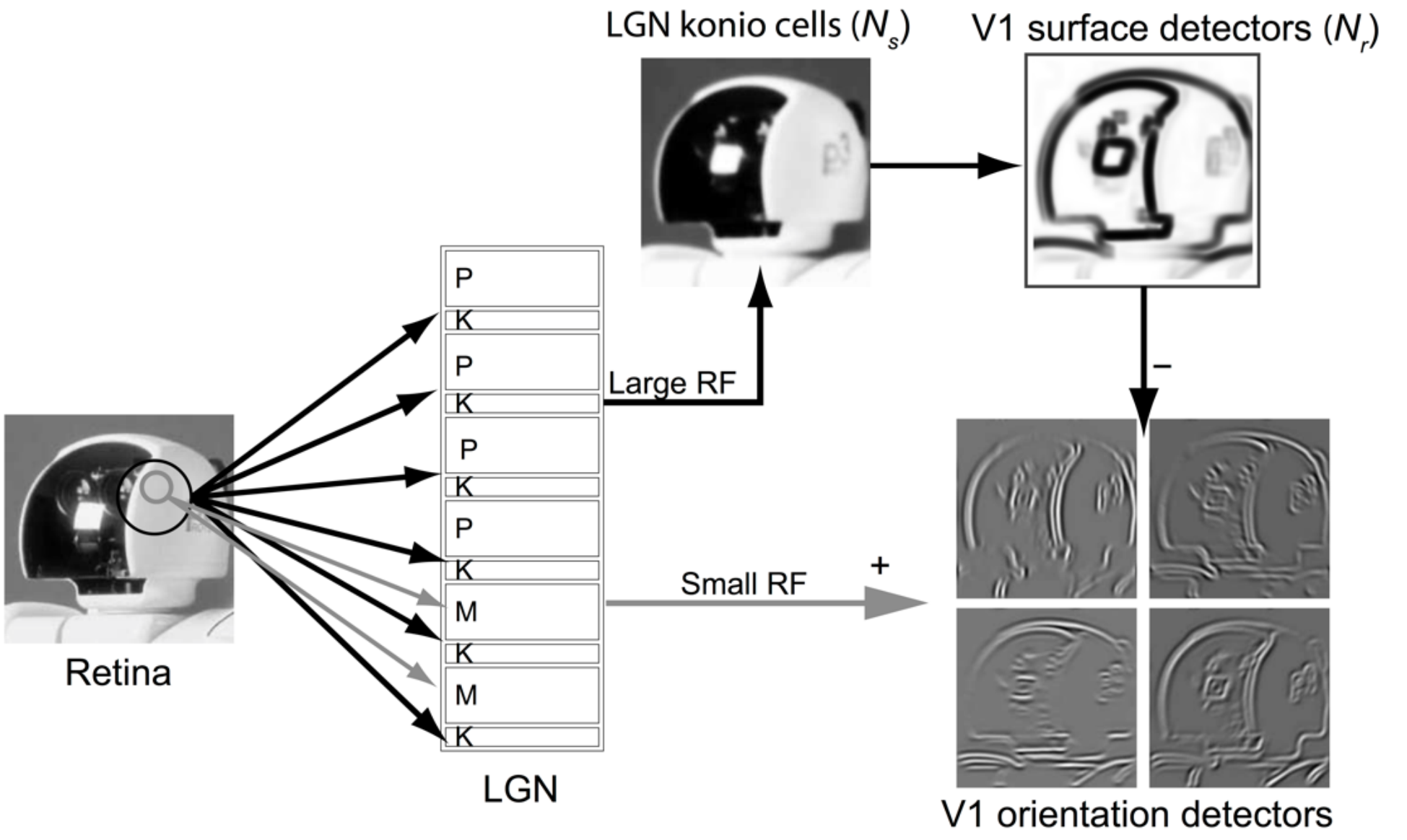

**Figure 3:** Illustration of the V1 model. Surface-selective neurons inhibit orientation-selective cells, suppressing their responses in homogeneous image areas. For details, see main text.

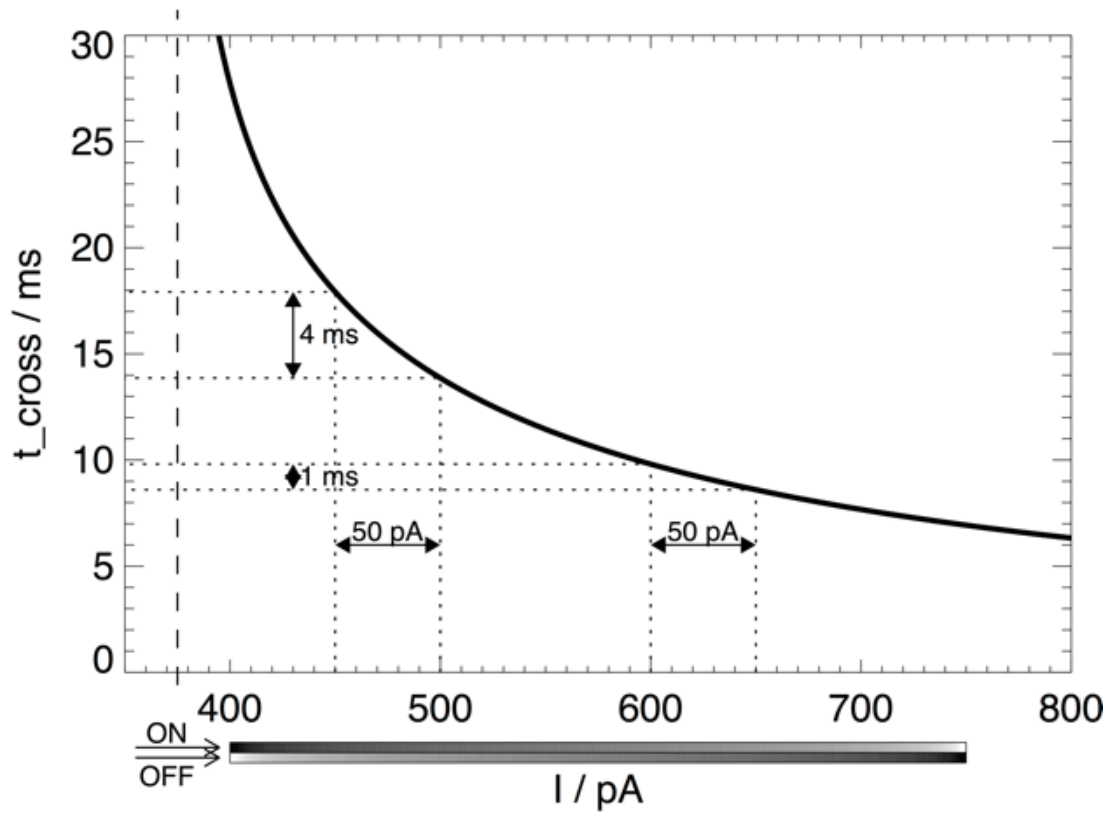

**Figure 4:** Current/latency relation (Eq. 9) for the injection of a constant current $I$ into a leaky integrate and-fire model neuron ($\tau_m = 10ms$, $E_l = -70\ mV$, $V_{th} = -55\ mV$, $R_m = 40\ M\Omega$). $t_{cross}$ denotes the time of the first spike after stimulation onset. The grey scales below the abscissa indicate how post-retinal activation maps to injection currents of 400– 750 $pA$. The sigmoidal transfer function of LGN konio-cells is mirror-symmetric for ON and OFF-cells. Current is mapped nonlinearly; a range of 50 $pA$ maps to different latency ranges depending on the absolute amplitude (*dotted lines*). This motivates the use of ON- and OFF-channels, to better resolve small differences in pixel luminance towards the extreme ends of the current spectrum (Sec. 2.1.1).

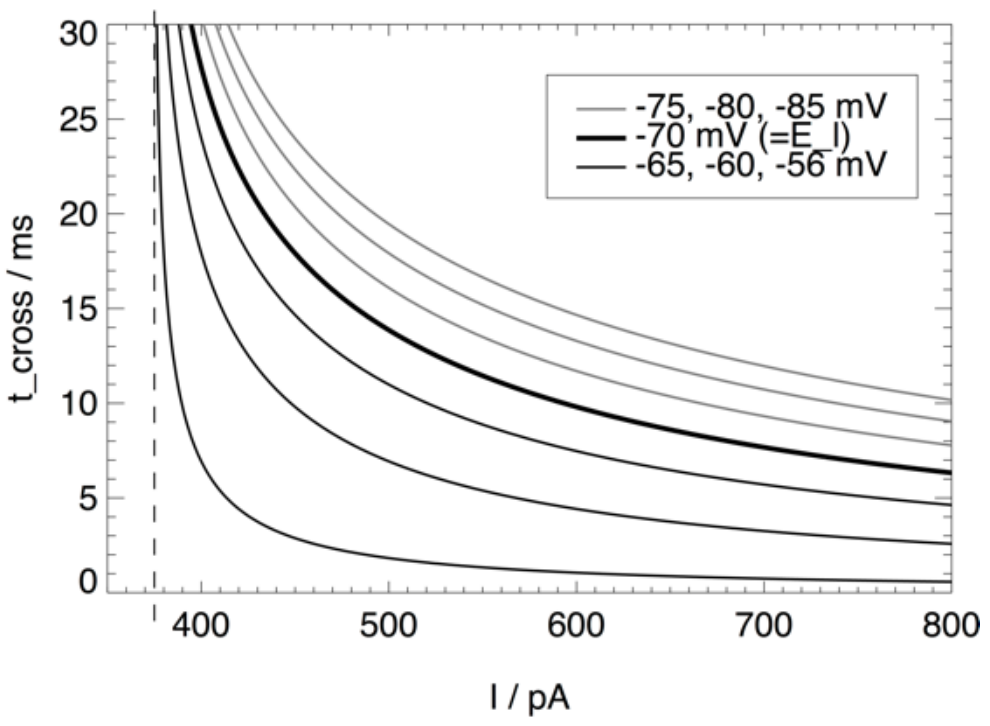

**Figure 5:** Current-latency relation (Eq. 8) for different starting membrane potentials $V_{start}$. *Thick curve,* starting from resting state ($V_{start} = E_l$, corresponds to Fig. **4**); *thin curves,* starting from depolarized states ($V_{start} < E_l$); *grey curves,* starting from hyperpolarized states ($V_{start} < E_l$); *dashed line*, $I$ = 385 *pA*, at which no spike is generated.

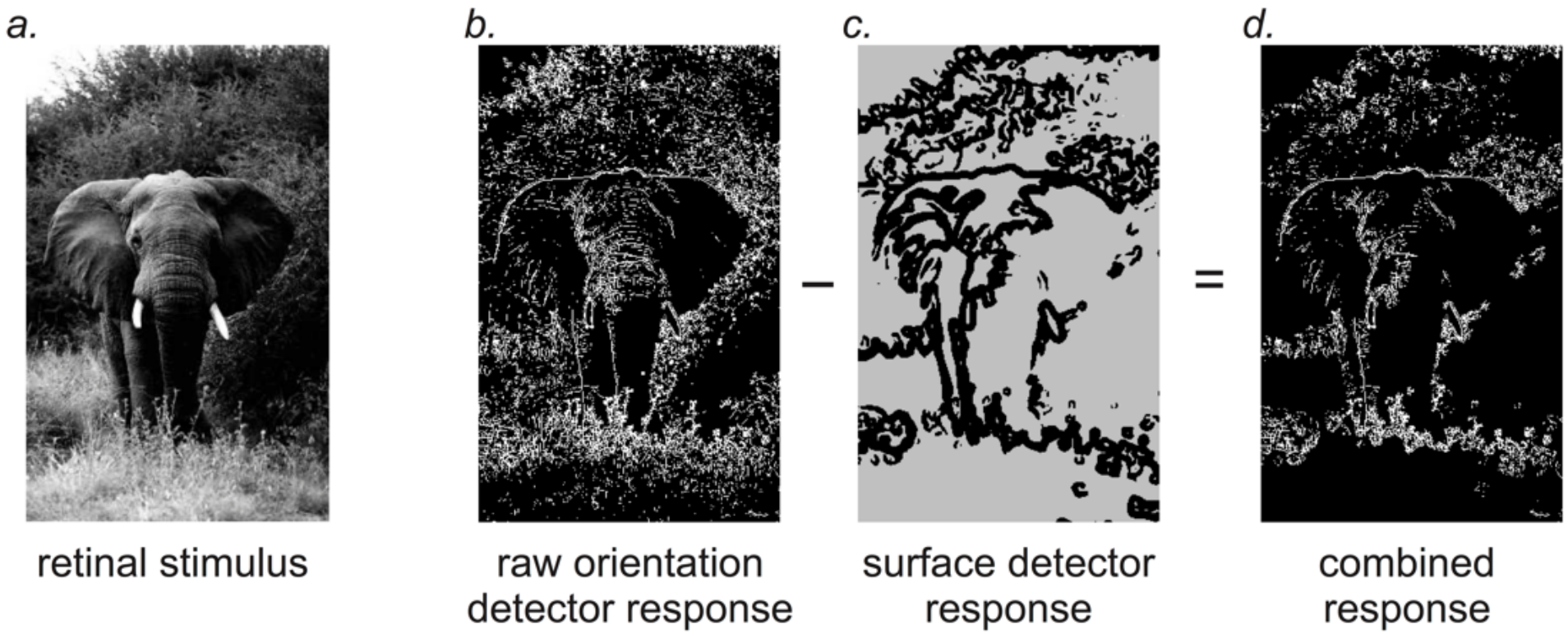

**Figure 6:** Simulation result: surface-selective neurons inhibited orientation-selective neurons. *a,* stimulus; *b,* response of orientation-selective neurons *without* inhibition (all orientations shown, bright pixels indicate neurons that produced at least one action potential); *c,* response of surface-selective neurons (ON and OFF); *d,* response of orientation-selective neurons *with* inhibition by surface-selective cells: responses on potential objects are reduced (compare to response in *b*).

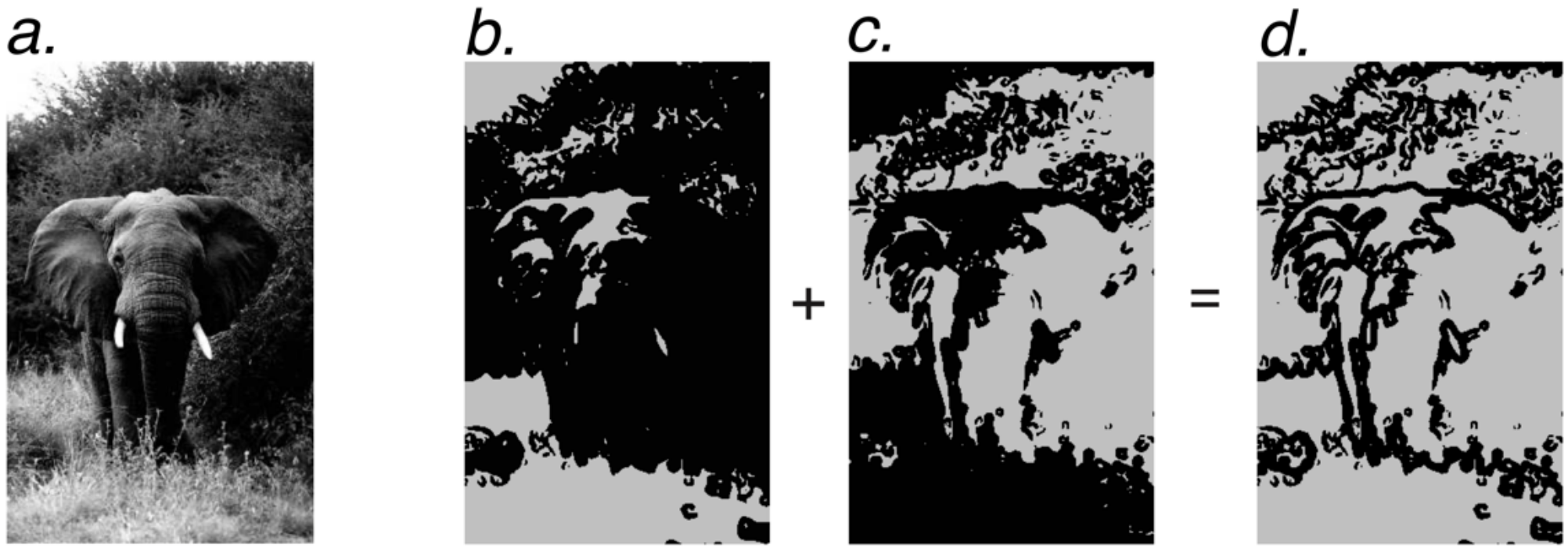

**Figure 7:** Complementary responses of surface-selective ON- and OFF-neurons. A) the retinal stimulus, b) homogeneity ON response, b) homogeneity OFF response, C) superposition of ON and OFF.

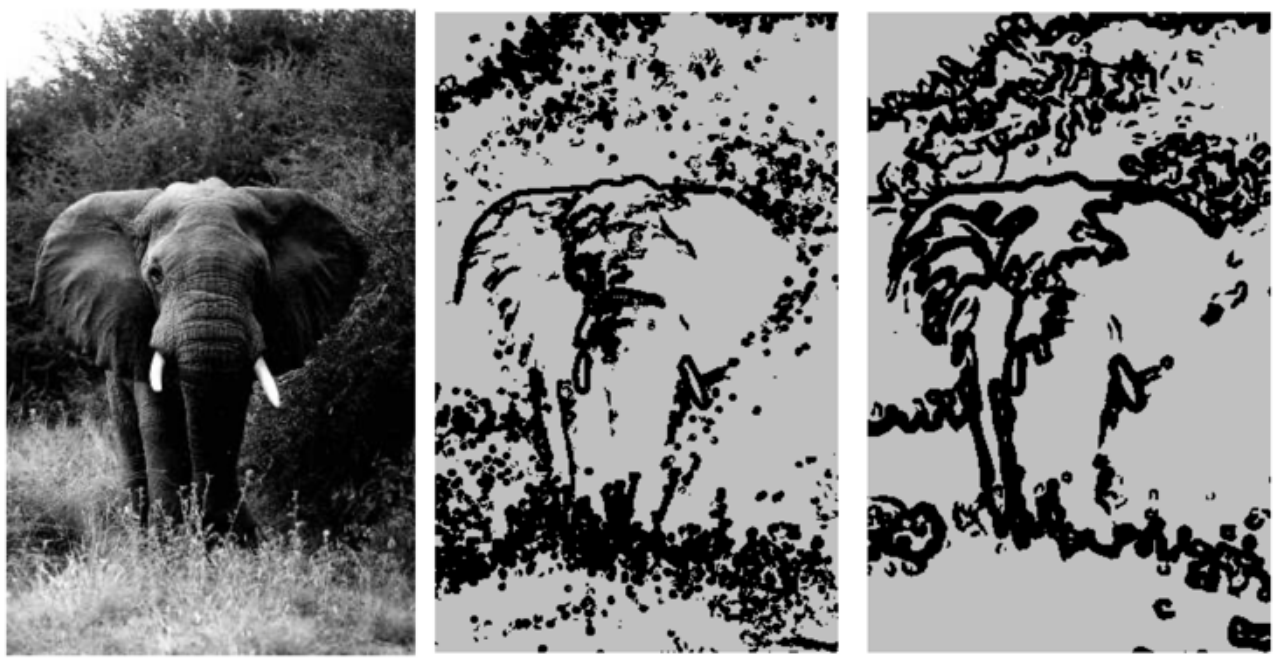

**Figure 8:** Effect of spatial low-pass filtering by large LGN receptive fields. The focus of surface detection shifts toward mid- or large-sized structures in the visual field. The effect can be well observed in the “elephant”-image, which contains many small structures that are not attributed to objects of a relevant physical scale.

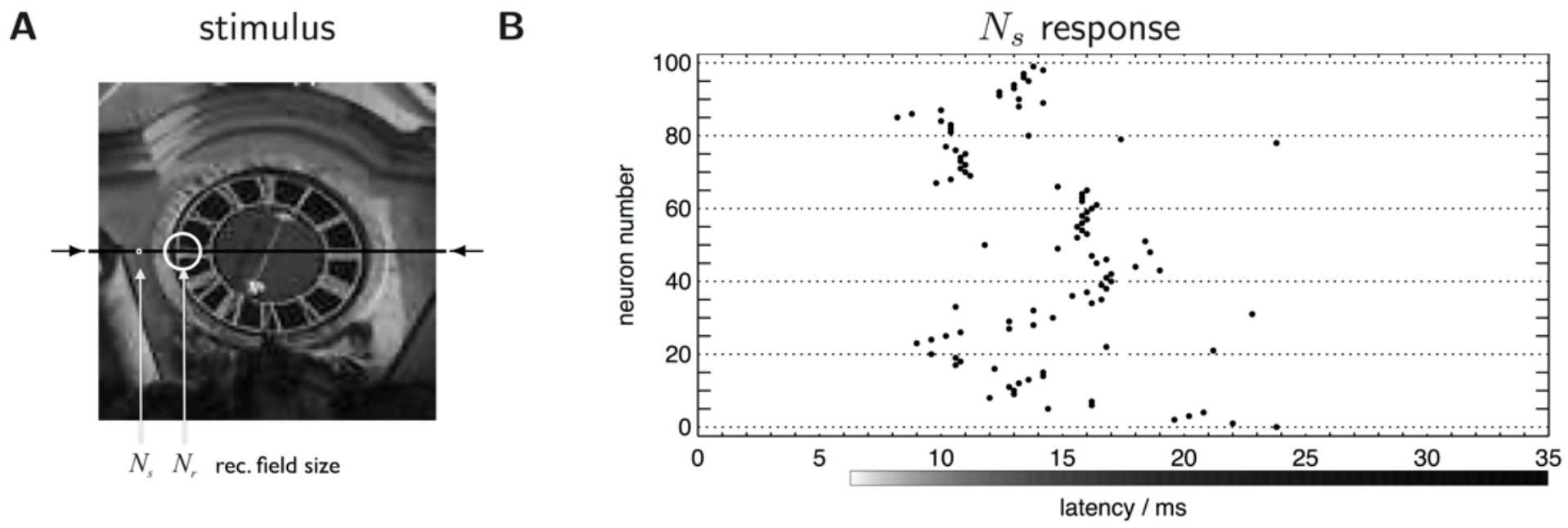

**Figure 9:** *A,* visual stimulus patch; *small circle,* receptive field of an $N_s$ neuron; *big circle,* receptive field of an $N_r$ neuron. *B,* typical single-trial spike trains of 100 $N_s$ neurons with receptive field positions along the marked line in *A*; *gray scale on latency axis,* corresponding pixel luminance by (Eqs. 10–12). Clocktower image courtesy of Antonello Ceravola.

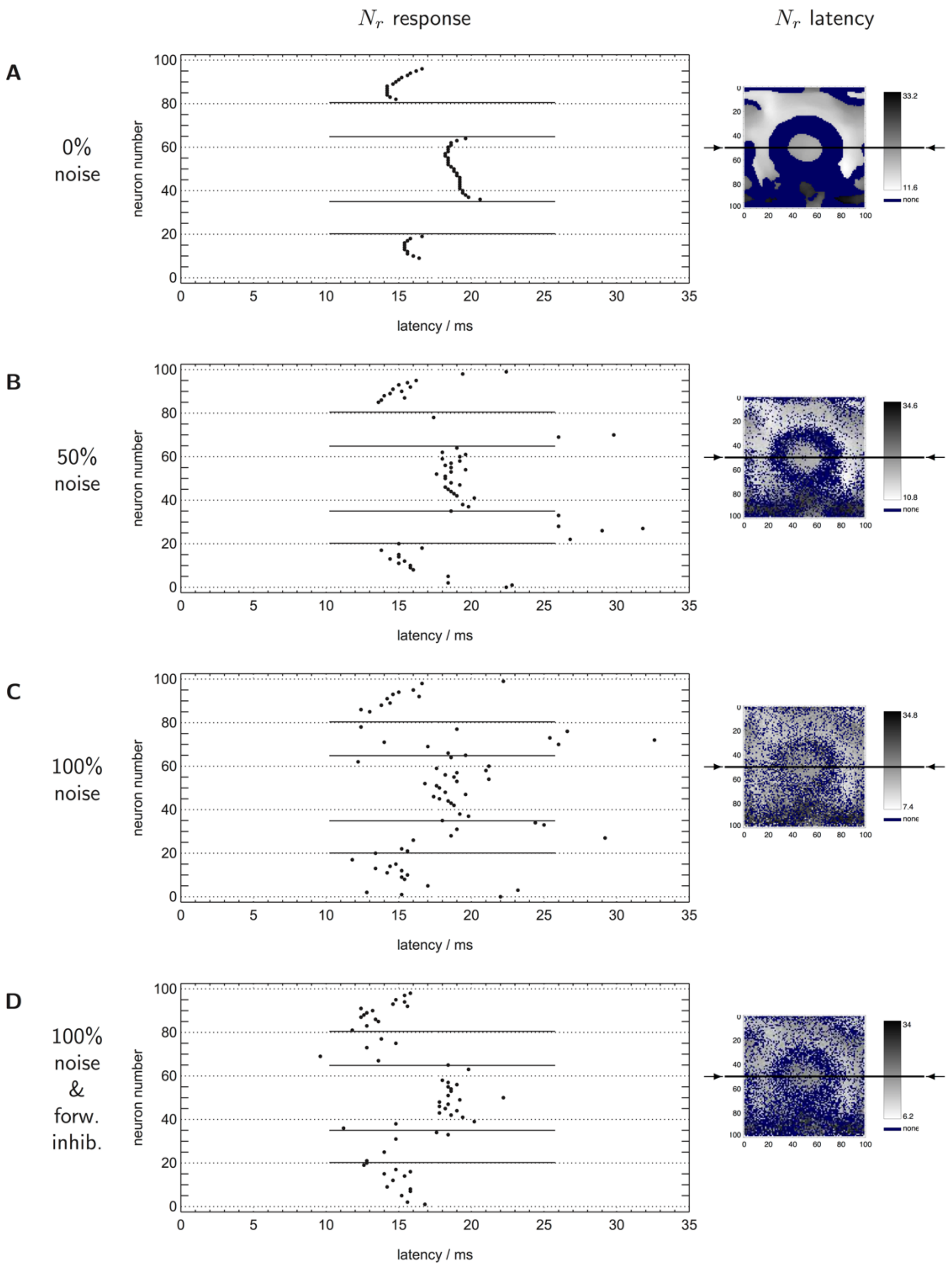

**Figure 10:** Typical single-trial homogeneity responses of 100 $N_r$ neurons with receptive field positions along the marked line in Fig. **9***A*. *Left,* spike train raster plots; *right,* latency map of all $N_r$ neurons. *A–C,* With increasing spike noise from the background population activity, spike response times scatter more. Moreover, the noise background triggered spikes in neurons which did not respond with zero noise input (blue *marked regions*). *D,* Additional feedforward inhibition counteracted the effect of background noise.

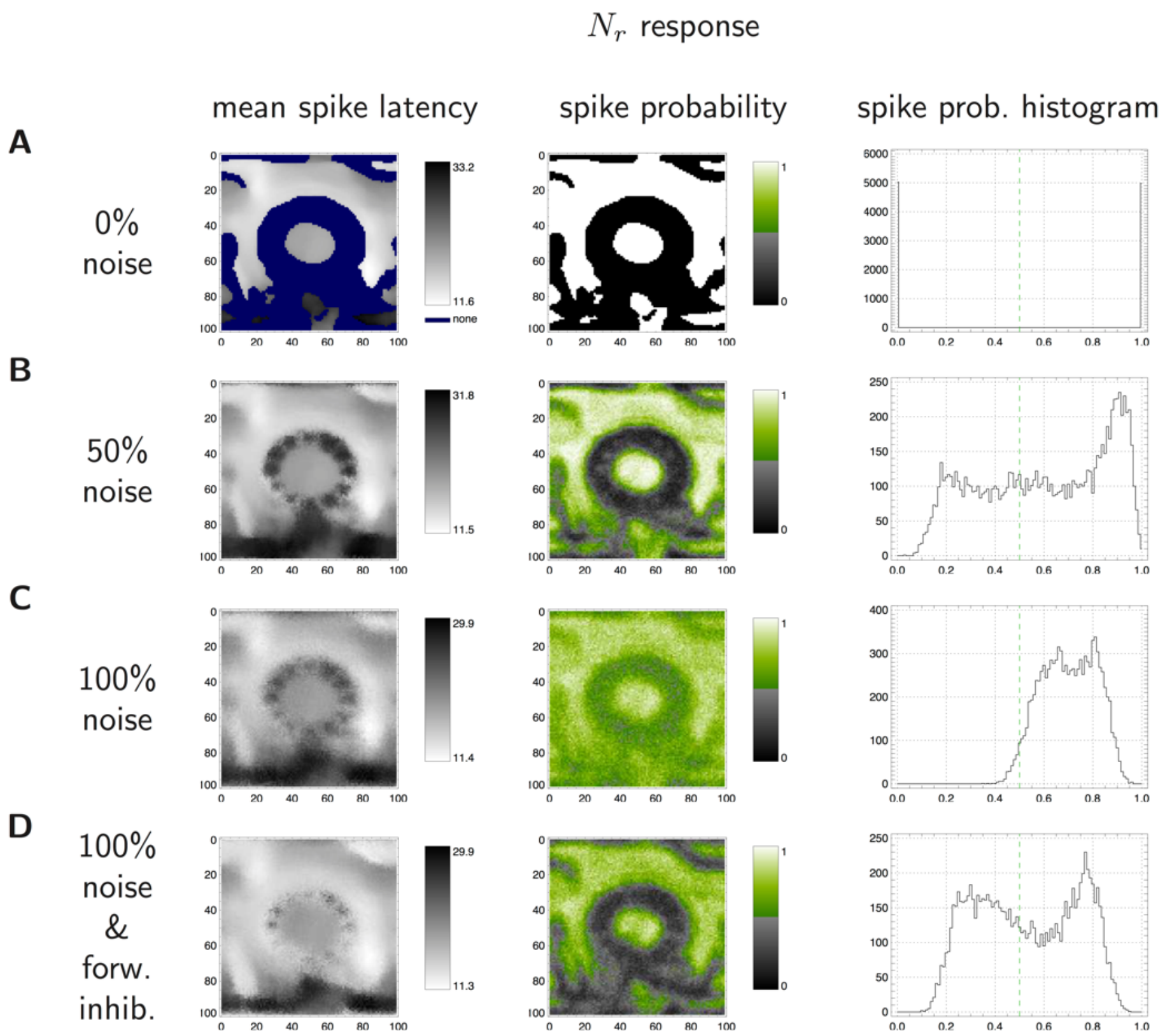

**Figure 11:** Mean $N_r$ spike latency, spike probability, and distribution of spike probabilities in ensembles of 100 neurons at each position in the view field. *A–C,* With increasing spike noise from the background population activity, the separation between homogeneous and non-homogeneous image regions breaks down; a level of 100% corresponded to the amount of crosstalk expected in the brain. *D,* Additional feedforward inhibition reestablishes reasonably good separation.

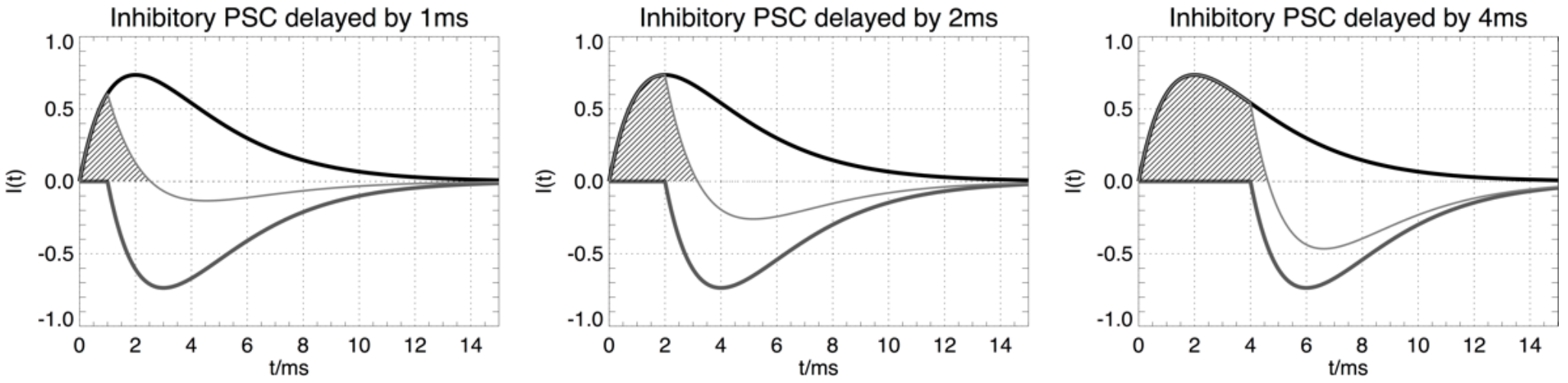

**Figure 12:** Shaping an excitatory PSC by delayed feedforward inhibition. The excitatory PSC is followed by an inhibitory PSC of the same shape and magnitude. The plots show superpositions for different delays of the inhibitory PSC ($\Delta t$ = 1, 2, or 4 ms). *Black curve,* excitatory PSC with time constant $\tau$= 2 ms; *grey curve,* inhibitory PSC with $\tau$= 2 ms and delay $\Delta t$; *thin curve,* effective PSC resulting from superposition. If spikes converge on a homogeneity-selective neuron, the positive parts of their resulting PSCs need to overlap in order to effectively depolarize the neuron to the spiking threshold. Note how the positive part of the effective PSC changes width with $\Delta t$: Shorter delays require spikes to arrive closer in time, i.e., they require increased homogeneity for the neuron to respond.

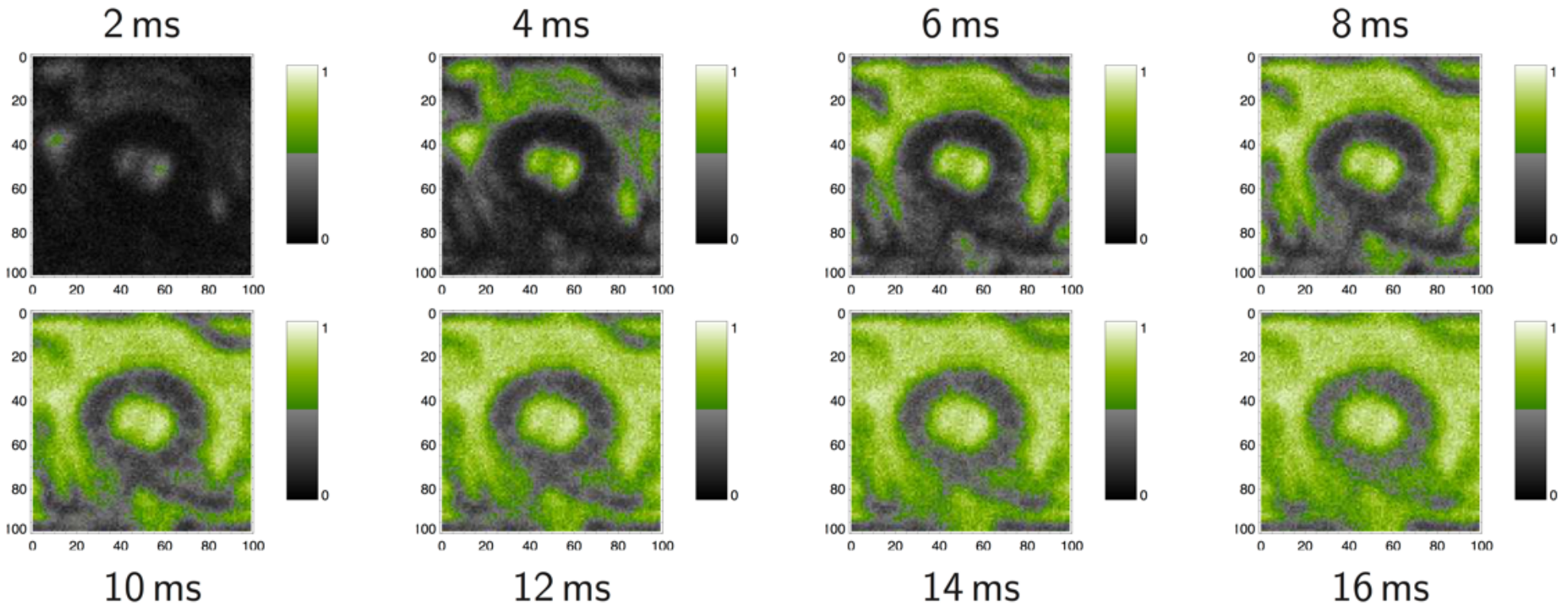

**Figure 13:** $N_r$ spike probability in 100 simulation runs for different feedforward inhibitory delays. Larger delays shifted the homogeneity threshold toward lower values (more regions were marked as homogeneous).